\documentclass[sigconf]{acmart}
\AtBeginDocument{%
  }

\setcopyright{none}
\acmDOI{}
\acmISBN{}
\acmConference[LOCO '26]{2nd International Workshop on Low Carbon Computing}{September 10--11, 2026}

\usepackage{hyperref}
\usepackage{subcaption}

\begin{document}

\title{Scoring and Gamification to Encourage Sustainable Use of Compute Clusters}

\author{Maximilian MacDonald, Chris McCaig, Sean MacAvaney, Matthew Barr, Lauritz Thamsen}
\affiliation{%
  \institution{University of Glasgow, United Kingdom --- \{firstname.lastname\}@glasgow.ac.uk}
  \country{}
}

\renewcommand{\shortauthors}{MacDonald et al.}

\begin{abstract}
  The environmental cost of computing continues to grow, yet behaviour change remains limited. We present a composite sustainability score integrating average carbon intensity, resource utilisation, and embodied emissions into a single 0–100 metric designed for gamified feedback. Each component rewards a different dimension of sustainable behaviour: carbon-aware workload shifting, high resource utilisation, and selecting hardware that is commonly underutilised. This scoring system is built into an existing cluster management interface and underpins three dashboard conditions: raw metrics, composite score, and a gamified tree visualisation, which we are planning to evaluate in a 12-week within-subjects study with approximately 35 researchers. Furthermore, we open the discussion on the challenge of defining computational work `goodness' in the context of sustainability scores.
\end{abstract}

\keywords{Low Carbon Computing, Behaviour Change, Persuasive Systems, Operational Emissions, Embodied Carbon, Cluster Computing}

\maketitle

\section{Introduction}
The environmental impact of the computing sector is a growing concern, with its emissions now rivalling industries such as aviation ~\cite{freitag_real_2021}.
Cluster and cloud computing energy use continues to grow at alarming rates, with data centres accounting for around 180 million tons of $CO_2$ per year from energy consumption alone and projected to grow to over 300 million tons by 2030\footnote{\url{https://www.iea.org/reports/energy-and-ai}}. 
Machine Learning (ML) in particular has come under scrutiny for its resource consumption, leading to calls for systematic reporting of energy and carbon metrics ~\cite{henderson_towards_2020}.

While awareness of computing's environmental impact is growing ~\cite{salam_challenges_2018}, awareness alone appears insufficient to drive meaningful behaviour change ~\cite{kumar_bagla_awareness_2022}. 
Many computing professionals are unaware of mitigation strategies~\cite{salam_challenges_2018}, and even when informed, cost, time, and a lack of accepted standards limit behaviour change~\cite{Sriraman2023AST}.
This awareness-action gap represents a critical barrier to achieving sustainability goals in computing, and so tools to promote behaviour change are needed.

Current approaches to computational sustainability focus primarily on measurement and reporting. Tools like CodeCarbon ~\cite{lottick_energy_2019} and the Green Algorithms calculator ~\cite{lannelongue_green_2021} enable researchers to estimate carbon footprints of individual jobs, while cloud providers have introduced carbon dashboards showing aggregate emissions ~\footnote{\url{https://docs.cloud.google.com/carbon-footprint/docs/methodology}} ~\footnote{\url{https://learn.microsoft.com/en-us/power-bi/connect-data/service-connect-to-emissions-impact-dashboard}}.
Meanwhile, gamification techniques, including feedback mechanisms, goal-setting, and social comparison, have demonstrated success in related domains, with residential energy consumption studies achieving up to 20\% reductions ~\cite{iria_gamification_2020}. 

Standard energy and carbon metrics lack intuitive meaning and fail to provide actionable guidance—users see numbers but do not know what constitutes "good" or how to improve, a common problem with purely numerical metrics ~\cite{larrick_economics_2008}. Single metrics like Carbon Intensity, while useful for specific optimisations such as temporal workload shifting, fail to capture the multidimensional nature of sustainable computing, which includes energy efficiency, embodied carbon, and total lifecycle use, among others. Optimising for one dimension such as embodied emissions can lead to inefficiency in another like operational emissions, highlighting the need for approaches that balance competing sustainability goals rather than optimising metrics ~\cite{hanafy_war_2023,acun_carbon_2023}.
Additionally, gamification has not been applied to behaviour change in sustainability computing, a promising domain where users are technically proficient and receptive to optimisation challenges when provided with clear goals.

We address these challenges through a composite sustainability score that integrates multiple dimensions: average carbon intensity, resource utilisation, and hardware lifecycle considerations. We combine this into a single 0-100 scoring metric suitable for gamified feedback. We present our design for a 12-week exploratory study on a mid-sized ML research cluster with approximately 35 active users to test the effectiveness of this scoring metric. Participants rotate through three interface conditions: raw metrics, composite score, and a gamified tree visualisation. We aim to measure both attitudinal changes and behavioural outcomes to evaluate whether gamified sustainability feedback drives more sustainable computing practices.

\textbf{Contributions.} We present the following contributions:
\begin{itemize}
    \item We design a gamifiable composite sustainability score that integrates average carbon intensity, resource utilisation, and a hardware bonus from underutilised resources, enabling multidimensional feedback suitable for gamification.
    \item We present our design for an 12-week study on a mid-sized research compute cluster that focuses on ML research.
    \item We open the discussion on evaluation of computational work ``goodness'', and how this challenge goes beyond resource efficiency.
\end{itemize}
\section{Related Work}
\subsection{Current Sustainability Metrics}
Beyond energy consumption and raw carbon usage, metrics have been developed to capture the various dimensions of computational sustainability. %
Carbon Intensity is the most widely adopted of these and has shown its usefulness through temporal workload shifting, enabling users to schedule jobs during periods of low grid intensity ~\cite{wiesner_lets_2021} with tools like CASPER ~\cite{souza_casper_2024} reducing the carbon footprint of workloads by up to 70\%.
However, Carbon Intensity ignores resource utilisation efficiency, embodied carbon, and hardware lifecycle considerations, providing little actionable guidance for users with deadline-driven work who cannot defer jobs.
Embodied carbon, the scope 3 emissions associated with hardware manufacturing, represents another critical dimension. Standard accounting methods spread manufacturing emissions over projected hardware lifespans, with \citet{zhang_spatial-temporal_2024} demonstrating significant variation in embodied carbon calculations depending on spatial-temporal modelling assumptions. \citet{gupta_chasing_2022} quantified that hardware production often equals or exceeds lifetime operational emissions, establishing embodied carbon as a sustainability lever comparable in importance to operational efficiency. 
Resource utilisation likewise influences sustainability outcomes. \citet{barroso_energy_2019} highlight that computing hardware exhibits significant baseline power draw when at low utilisation. While advancements in energy proportionality have resulted in an almost linear relationship after the initial idle cost, higher utilisation still spreads both operational energy and embodied carbon across more useful work, reducing per-task environmental impact.
Some tools have made use of multidimensional sustainability metrics. For example \citet{acun_carbon_2023} developed Carbon Explorer, a framework integrating operational carbon, embodied carbon, and hardware lifecycle considerations for datacenter design. Sustainability dashboards, such as the Green Algorithms Dashboard~\cite{lannelongue_green_2021} and Kepler~\cite{amaral_kepler_2023}, similarly focus on communicating energy and carbon costs, but offer little insight into actionable behaviour change, and are generally designed for cluster administrators rather than individual users. 

These multidimensional approaches reveal inherent tensions. \citet{hanafy_war_2023} demonstrated that optimising for carbon efficiency can lead to energy inefficiency, while \citet{lee_carbon_2024} showed that as energy grids become cleaner, the relative importance of embodied carbon increases. This remains the case even when older hardware consumes more energy per computation, and particularly when continuing to use it avoids the need to acquire new hardware. These competing priorities highlight the challenge of designing single metrics that capture sustainability's multidimensional nature.
Additionally, no prior work has combined these dimensions into a single metric suitable for guiding individual job submission decisions.%
\subsection{Gamification for Sustainability}

Gamification, defined as adding game-like elements to non-game contexts ~\cite{deterding_game_2011}, has been applied to residential energy consumption, transportation, and waste reduction to positive effects. \citet{iria_gamification_2020} achieved energy reductions through dashboards combining real-time feedback with social comparison, while a gamified garden representation focusing on aesthetic feedback reduced user energy consumption by 0.5 standard deviations in an office context ~\cite{oppong-tawiah_developing_2020}

\citet{koivisto_rise_2019} conducted a systematic review of 800+ gamification studies, finding that successful interventions combine extrinsic and intrinsic motivational elements, balance short-term rewards with long-term goals, and include mechanics that encourage real-world action.
Meanwhile, \citet{nicholson_recipe_2015} defines Meaningful Gamification and highlights the importance of designing the gamified system with the user at its centre and allowing them to engage with the system on their own terms. This builds on ideas of autonomy, competence, and relatedness presented in Self Determination Theory to drive behaviour change ~\cite{ryan_self-determination_2000}.
Scoring systems form a core component of gamification, providing clear targets for improvement and enabling social comparison. However, \citet{hanus_assessing_2015} demonstrated that poorly designed gamification can backfire, undermining intrinsic motivation. In sustainability contexts specifically, metrics must balance intrinsic motivations, such as environmental concern, with extrinsic rewards, like points ~\cite{mekler_towards_2017}. 
Poorly designed systems can encourage gaming behaviours, where users appear sustainable without genuine change, or create inequitable competition disadvantaging users with legitimate high-resource needs.
Related work on eco-feedback technology provides design principles for sustainability interfaces. \citet{froehlich_design_2010} established that effective eco-feedback requires granular data, immediate feedback, and clear action-outcome relationships. \citet{wu_climate_2015} found that interactivity and visualisation enhance learning and promote sustained behaviour change beyond passive information exposure. 
Despite the solid body of work on gamification for promoting sustainable behaviour no work has applied these concepts to a computing context, or applied scoring methods to sustainability computing metrics.

\section{Ongoing Work}

In this section we present the design of an upcoming 12-week deployment study exploring the effect of exposing researchers to multi-dimensional sustainability feedback, and whether this can encourage more sustainable resource usage on a mid-sized ML research cluster.
We then present both the composite sustainability scoring metric underpinning the intervention and the gamification techniques through which it is communicated to users.
As the deployment is scheduled to begin in the coming weeks, this section focuses on the design, implementation details, and planned evaluation methodology underlying the intervention.

\subsection{Study Design}

The study is embedded within the daily workflows of research cluster users at the University of Glasgow School of Computing Science and employs a mixed-methods, sequential explanatory design with approximately 35 active users, in which quantitative behavioural data inform subsequent qualitative interviews. The intervention is integrated into Launcher, the custom web application cluster members use to monitor resources and deploy containerised jobs via an OKD Kubernetes interface. This integration ensures access to users' natural work environment without requiring adoption of new tools. Operational carbon emissions are calculated dynamically using real and estimated hardware power alongside real-time regional grid intensity data from the UK National Grid Carbon Intensity API \footnote{\url{https://www.carbonintensity.org.uk/}}. GPU power consumption is measured directly via NVIDIA System Management Interface, while CPU power is estimated using a linear model to Thermal Design Power (TDP) based on utilisation. 
While 0-to-TDP power models are not necessarily accurate for low resource utilisation, given static power draw, they are widely available, including for our heterogeneous hardware configurations. Resource utilisation of CPU, GPU, and memory is measured directly from cluster nodes at 5-minute intervals.

\begin{figure*}
    \centering
    \begin{subfigure}{0.45\textwidth}
        \centering
        \includegraphics[width=\textwidth]{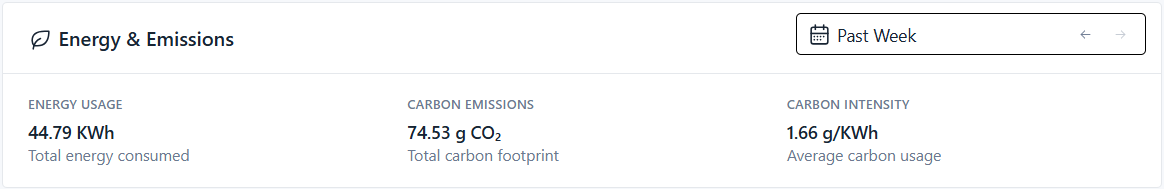}
        \caption{Raw metrics view}
        \label{fig:RawMetric}
    \end{subfigure}
    \hfill
    \begin{subfigure}{0.5\textwidth}
        \centering
        \includegraphics[width=\textwidth]{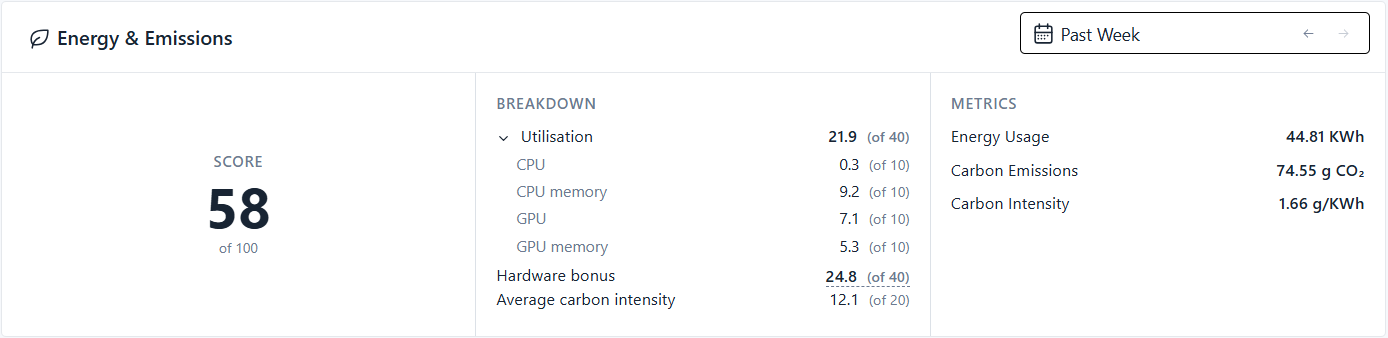}
        \caption{Composite sustainability score view}
        \label{fig:Score}
    \end{subfigure}

    \vspace{1em}
    \begin{subfigure}{\textwidth}
        \centering
        \includegraphics[width=.95\textwidth]{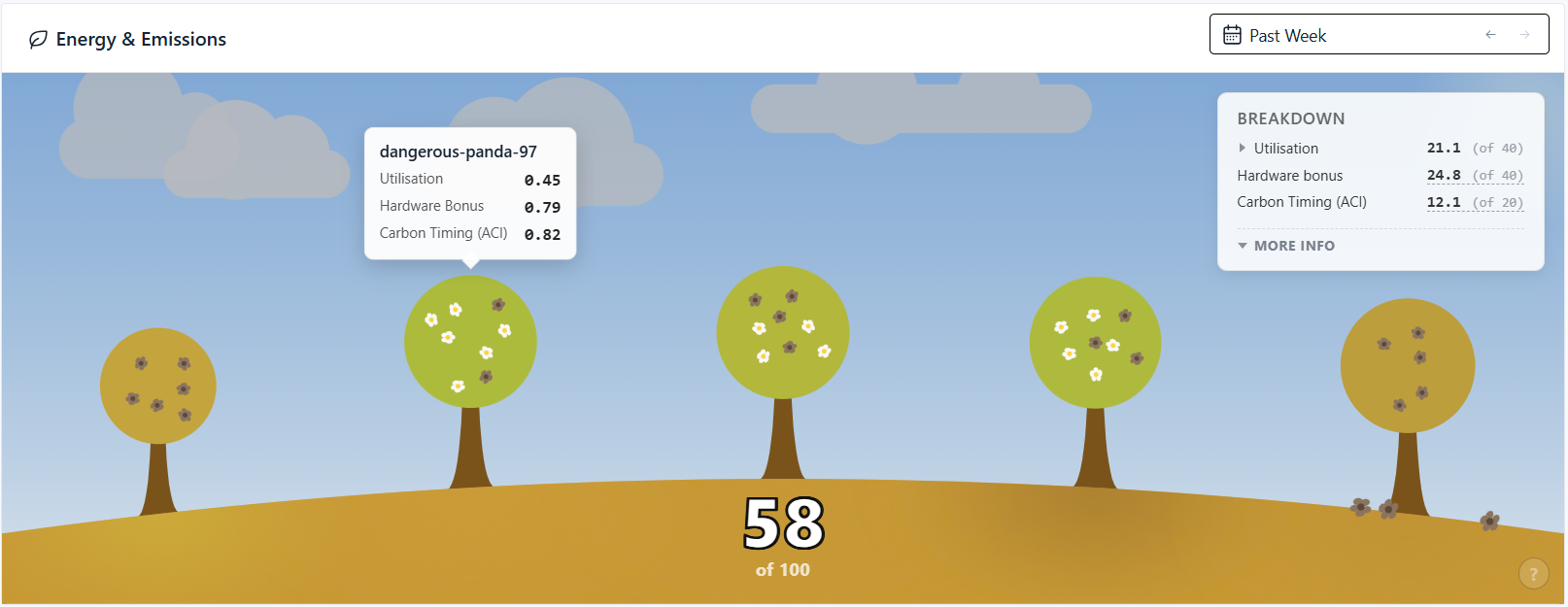}
        \caption{Gamified view}
        \label{fig:Gamified}
    \end{subfigure}
    
    \caption{Screenshots of the three different dashboard configurations}
    \label{fig:confusion_matrices}
\end{figure*}

\subsubsection{Dashboard Visualisation Configurations}
Participants are exposed to three distinct dashboard configurations. Participants rotate through these dashboard configurations in a randomised order, each presented for 3 weeks to control for temporal and learning effects, alongside 3 weeks of a control with no change to the dashboard.
\begin{enumerate}
    \item \textbf{Raw Metrics Baseline (Fig.\ref{fig:RawMetric}):} This configuration integrates fundamental, quantitative metrics directly into the user's dashboard. It displays raw energy consumption in kilowatt-hours, carbon footprint in grams of CO$_2$ equivalent, and the average carbon intensity over a period.
    \item \textbf{Composite Sustainability Score (Fig.\ref{fig:Score}):} This view condenses multi-dimensional sustainability data into a single score between 0 and 100. The score balances average carbon intensity with efficient hardware utilisation across CPU, GPU, and memory allocations, whilst offering a ``hardware bonus'' to incentivise underutilised infrastructure. Breakdowns of the score are supplied alongside the overall score.
    \item \textbf{Gamified Representation (Fig.\ref{fig:Gamified}):} This configuration aims to apply interface aesthetics and narrative themes to communicate ecological costs and provide feedback. The composite score is mapped to an interactive forest visualisation, where trees represent specific pods, with the colour and liveliness of each corresponding to hardware bonus. The weather of the background reflects the users average carbon intensity, while ground colour and gradients from each tree show utilisation contributions. Dynamic elements aim to reinforce user relatedness, and tooltips provide access to underlying pod contribution metrics for more detailed inspection.
\end{enumerate}

\subsubsection{Surveys and Qualitative Interviews}

Upon consent via in-dashboard recruitment, participants will complete a baseline survey measuring awareness of sustainable computing concepts, attitudes toward sustainability responsibility, current practices, and perceived barriers. Participants are then randomly assigned to one of six condition sequences to counterbalance order effects. Post-intervention surveys repeat key measures and add questions on score interpretation and condition preferences. Throughout the study, job submission patterns, resource allocation, utilisation, hardware selection, and dashboard engagement are continuously logged to characterise behavioural trends across conditions, providing exploratory context for the qualitative interviews. These post-study semi-structured interviews with 10--15 participants will be analysed using reflexive thematic analysis~\cite{braun_reflecting_2019} to explore score interpretation, decision making, and barriers to sustainable practice.

\subsection{Composite Sustainability Score Design} \label{ScoreDesign}

We designed a gamifiable composite sustainability score to provide multidimensional feedback suitable for gamification. Each component is normalised to a 0--1 scale and weighted.

\subsubsection{Resource Utilisation}

We measure utilisation across four dimensions: GPU usage, CPU usage, GPU memory, and CPU memory. For each job, we calculate the average utilisation of each resource type over its reserved duration. 
Each utilisation dimension is scored linearly, with 0\% utilisation receiving a score of 0 and 100\% utilisation receiving a score of 1. In cases of over-utilisation, where actual usage exceeds requested resources, we cap the score at 1 rather than penalising users. While this creates potential for gaming the system through deliberate under-allocation or artificial over-utilisation, we assume good faith and that resources were genuinely needed.
The overall score is the average of the four resource dimensions:

\begin{equation}
U_{\text{score}} = \frac{1}{4}\left( u_{\text{GPU}} + u_{\text{CPU}} + u_{\text{GPU-mem}} + u_{\text{CPU-mem}} \right)
\end{equation}
where each component $u_r = \min(1.0, (\text{actual}_r / \text{allocated}_r))$ represents the utilisation efficiency for resource $r$.

\subsubsection{Hardware Bonus}

Our hardware bonus incentivises use of underutilised resources, often older or less popular hardware configurations. We calculate the average utilisation of each GPU type across the cluster over a rolling one-week window. The hardware bonus awards points inversely proportional to recent utilisation: the least-used GPU type receives the maximum bonus, while fully utilised GPUs receive no bonus. Specifically, for a job using GPU type $g$:

\begin{equation}
H_{\text{bonus}}(g) = \frac{1 - U_g}{1 - U_{\text{min}}}
\end{equation}
where $U_g$ is the cluster-wide utilisation of GPU type $g$ over the past week and $U_{\text{min}}$ is the utilisation of the least-used GPU type. This ensures that the greatest incentives exists for the most underutilised resources while maintaining some reward for high but not fully utilised hardware.

\subsubsection{Average Carbon Intensity (ACI)}

Our ACI component evaluates how well a job's timing aligns with available low-carbon periods within a realistic scheduling window. For each pod, we compare its actual ACI to the theoretically optimal intensity achievable within a constrained window. The scheduling window is limited to one week or twice the job's duration, whichever is highest.
For a job of duration $n$ timestamps, we identify the optimal scheduling window by finding the contiguous block of at least $n$ consecutive timestamps with the lowest average carbon intensity, and the worst-case block as the contiguous sequence with the highest. The actual ACI is then mapped linearly between these bounds:

\begin{equation}
\text{ACI}_{\text{score}} = 1 - \frac{\text{ACI}_{\text{actual}} - \text{ACI}_{\text{optimal}}}{\text{ACI}_{\text{worst}} - \text{ACI}_{\text{optimal}}}
\end{equation}
A score of 1 represents optimal timing and 0 represents the worst possible timing. As such users score highly by actively choosing to run their workloads at times with relatively low ACI. To aggregate across multiple jobs at the user or namespace level, we compute a time-weighted average.

\subsubsection{Composite Score}

The final sustainability score combines and weights these components:

\begin{equation}
S_{\text{total}} = w_U \cdot U_{\text{score}}  + w_H \cdot H_{\text{bonus}} + w_{\text{ACI}} \cdot \text{ACI}_{\text{score}}
\end{equation}
where $w_U$, $w_{\text{ACI}}$ and $w_H$ are weights satisfying $w_U + w_H + w_{\text{ACI}} = 1$. Initial weights are set to $w_U = 0.4$, $w_H = 0.4$, and $w_{\text{ACI}} = 0.2$, reflecting that users have the most direct control over resource utilisation and hardware selection, and least control over grid carbon intensity. We expect to refine both the scoring mechanisms and gamification methods discussed below based on the feedback and insights from the upcoming study.

\subsection{Gamification Methods}

Making use of this composite scoring metric, the gamified dashboard view aims to incorporate gamification through feedback, aesthetic design, and lightweight narrative elements. 
Rather than presenting sustainability metrics in isolation, user actions will be reflected through evolving visual states within the interface. This approach aims to create a stronger connection between user behaviour and system response. By linking visual progression directly to user action, the system seeks to support relatedness and intrinsic motivation.
The visual representation takes the form of an interactive forest, whose appearance encodes multiple dimensions of the sustainability score simultaneously. Each tree represents a single pod within the selected time frame, with the colour of each tree's leaves corresponding to the hardware bonus of the pod, going from a vibrant green when healthy to brown/red when the hardware bonus is low. Additionally, each tree has flowers that first die, then fall from the tree as the hardware bonus decreases. Utilisation is represented by the ground colour, with the average utilisation over the time frame shown as the average ground colour and individual pod contributions shown through a gradient around the base of each tree.
The weather of the background changes to reflect the average carbon intensity, going from sunny and clear, to overcast, to stormy with a red glow on the horizon as the score declines. The size of each tree shows the individual ACI contributions of each pod.
The visual language is intentionally aligned with the sustainability domain, using environmentally-associated imagery and progressive visual states to reinforce the meaning of user actions through thematic congruence. Dynamic elements are intended to contribute towards a sense of immersion: leaves bob when hovered and clicked, clouds float across the scene with speed relative to the current weather, and fallen flowers can be cleared away by the user.

\section{Open Challenges}
Our current scoring method evaluates workloads by their carbon efficiency, resource utilisation, and hardware choice, assuming that the work being run is always valuable. However, the problem of defining 'good work' remains an open challenge. Self-evidently, the most sustainable option is to avoid running work that is not useful, for example work that leads to no significant improvement. However, determining whether work is indeed useless is difficult, particularly in a research context where trial and error and experimentation are fundamental.
One approach to quantifying useful or good work could be to encourage user reflection through self-evaluation. For example, asking users to rate the usefulness of a job after it completes, which may also prompt more deliberate consideration before launching it.
Conversely, care must be taken not to discourage potentially useful work. For example, running large ML tasks with minor parameter tweaks when model performance is already comparably high is arguably wasteful, but the line between exploration and inefficiency is not always clear. Perhaps the more productive framing is to encourage mindful experimentation, rather than attempting to penalise it.
More specifically there is the question of energy efficiency: how much energy is really needed to achieve a certain result? And, even if energy use is reduced, is the additional time spent to increase efficiency bought from even more unsustainable sources, e.g. carbon costs associated with housing, transport and consumable goods required by the user.
Even for factors that are directly measurable, like code efficiency, there are often multiple approaches to reach a result, with possibly different energy consumptions, but how optimal any specific computational load is, can be hard to capture.
Ultimately, the attempt to quantify the goodness of work may be inherently hard and, hence, of limited practicality, and behaviour change in this respect may need to be culturally motivated rather than metric-driven. This work does not attempt to resolve this question, but raises it as a necessary consideration for future work on our sustainability scoring system.

Failed jobs present a particular challenge. They consume energy and hardware lifetime without producing useful output, yet determining whether a failure stems from user mistakes, such as under-provisioning memory, or infrastructure factors like hardware faults, is non-trivial. Penalising users for failures outside of their control risks discouraging legitimate resource use, while ignoring failed jobs entirely removes incentive to submit well-configured workloads. As such developing reliable fault attribution remains an open challenge.

\section{Summary and Future Work}
This paper presents a gamified scoring mechanism to capture usage of cluster resources in a way that can be used in future gamification practices. It aims to reflect costs related to carbon efficiency, resource utilisation, and hardware choices, and in doing so open a pathway to incentivise sustainable computing behaviour. We show how this scoring system is composed and our design for an upcoming study to test the effect of this method on user attitudes and behaviour.

Future iterations of the gamified dashboard will incorporate additional gamification features which may include social and self-comparison mechanisms, where users compete against their own historical scores, collaborate with teammates to meet shared challenges, or compete with peers for higher rankings. We also aim to introduce longitudinal progress tracking to support long-term behaviour change, and to investigate how more explicit narrative structures affect user engagement. Finally, we wish to examine what effect genuinely engaging gamification has on user motivation and sustained behaviour change beyond informational feedback.

\bibliographystyle{ACM-Reference-Format}
\bibliography{references_filtered}

\end{document}